\documentclass{article} 
\usepackage{amsmath,amssymb,amsfonts}

\title{Photon rockets and
  the Robinson-Trautman geometries}

\author{Sergio Dain\thanks{Fellowship holder from CONICOR.} \and  Osvaldo M. Moreschi\thanks{Member of CONICET.} \and Reinaldo J. Gleiser\thanks{Member of CONICET.}\\
  Facultad de Matem\'atica Astronom\'{\i}a y F\'{\i}sica (FaMAF)\\
Universidad Nacional de C\'ordoba,\\
Ciudad Universitaria, \\
  (5000) C\'ordoba, Argentina}

\begin{document}
\maketitle

\begin{abstract}
  We point out the relation between the photon rocket spacetimes and
  the Robinson Trautman geometries. This allows a discussion of the
  issues related to the distinction between the gravitational and
  matter energy radiation that appear in these metrics in a more
  geometrical way, taking full advantage of their asymptotic
  properties at null infinity to separate the Weyl and Ricci
  radiations, and to clearly establish their gravitational energy
  content. We also give the exact solution for the generalized photon
  rockets. 

\end{abstract}

\section{Introduction}

Examples of spacetimes representing the so called photon rockets, have
been recently studied in the literature\cite{Bonnor94}\cite{Damour95}.
An important problem under consideration in those examples was the
existence of gravitational radiation associated with the accelerated
motion of the `rocket', and its relation with the outgoing energy
momentum, in the form of a `null (or photon) fluid', that accompanies
the accelerated rocket motion. In the example of reference
\cite{Bonnor94}, which is based on the Kinnersley\cite{Kinnersley69}
metric, the well-known result of the absence of gravitational
radiation is analyzed as an application of the quadrupole formula. A
problem in this type of analysis is the presence of singularities in
the metric, precisely along the `world-line' of the `accelerated'
object. To analyze this problem, and be able to assign a meaning to
concepts such as `rocket trajectory' and `distributional energy
momentum tensor', in reference \cite{Damour95} it is considered the
associated linearized problem, and its relation to the exact solution
of Einstein equation; obtaining conditions for the presence of a
non-vanishing flux of gravitational radiation.

In this note we present a different approach. We start by pointing 
out the relation between the photon rocket spacetimes and the 
Robinson-Trautman geometries. All these metrics are asymptotically 
flat. This allows us to present a more geometrical discussion 
of the issues related with the distinction between the gravitational 
and matter-energy radiation that appear in these metrics, taking 
full advantage of their asymptotic properties at future null 
infinity to separate the Weyl and Ricci radiations, and to clearly 
establish their gravitational energy content in the general case. 
We show in the last section that the exact solutions to the generalized 
photon rockets are the Robinson-Trautman geometries.

\section{The Robinson-Trautman geometries}

The vacuum solutions containing a congruence of diverging null 
geodesics, with vanishing shear and twist, were first studied 
by Robinson and Trautman\cite{Robinson62}. These metrics can be expressed\cite{Frittelli92} 
in terms of the following line element

\begin{equation} \label{pr1}
 ds^{2} =\left( -2Hr+K-2\frac{m(u)}{r} \right) du^{2}
+2\;du\;dr-\frac{r^{2} }{P^{2} } d\zeta \;d\bar{\zeta } 
\end{equation}
where $P=P(u,\zeta ,\bar{\zeta } )$, $H=\frac{\dot{P} }{P}$, $K=\Delta \ln P$, a doted quantity denotes its time derivative and $\Delta$ is the two-dimensional Laplacian for the two-surfaces $u=constant$, 
$r=constant$ with line element
\[
dS^{2} =\frac{1}{P^{2} } \;d\zeta \;d\bar{\zeta }.
\]

It is usually convenient to describe this line element in terms 
of the line element of the unit sphere; this is done by expressing 
$P$ in terms of 
 $P=V(u,\zeta ,\bar{\zeta } )P_{0} (\zeta ,\bar{\zeta } )$
, where 
 $P_{0} $
 is the value of $P$ for the unit sphere. 
If $l$ denotes the vector field that generates the null congruence, 
then  $l=du$,  $l(r)=1$, 
 $l(\zeta )=0$
 and 
 $l(\bar{\zeta } )=0$
. In other words this is the coordinate system 
adapted to the geometry.

In reference \cite{Robinson62} it was found that the vacuum Einstein equation 
can be reduced to a parabolic equation for a scalar depending 
on three variables, the so called Robinson-Trautman equation; 
which using the GHP\cite{Geroch73} notation has the form
\begin{equation} \label{pr2}
-3\;m\;\dot{V} =V^{4} \;\eth^{2}\bar \eth^2 V-V^3\eth^{2} V  \bar \eth^2 V;
\end{equation}
where we fixed the freedom of redefining $u$ in such a way that 
$m(u)$ is actually a constant, and 
 $\eth$
 is the GHP edth operator of 
the unit sphere. We will refer to a line element with $V$ satisfying 
this equation as a Robinson-Trautman solution. On the other hand, 
if the Robinson-Trautman equation is not required, then the solution 
is no longer vacuum and there is only one component of the Ricci 
tensor different from zero, given by
\begin{equation}\label{pr3}
\Phi _{22}^{(RT)} =\frac{-3\;m\frac{\dot{V} }{V}-V^3 \eth^2 \bar \eth^2 V +\eth^2 V \bar \eth^2 V }{r^2}  
;
\end{equation}
where the (RT) is to emphasize the fact that in this case we 
are using the null tetrad adapted to the null congruence. We 
refer to this as a Robinson-Trautman geometry.

Let us consider next some examples of theses geometries.

\section{First example: flat spacetime}

As an example, let us consider the case in which one requires 
a Robinson-Trautman geometry to be flat; which demands, among 
other things, $m$ to be zero. In this case the above equation describes 
the Minkowski line element in terms of a set of null polar coordinates 
associated with a timelike curve. The scalar 
 $V(u,\zeta ,\bar{\zeta } )$
, in this case, 
contains only 
 $Y_{00} (\zeta ,\bar{\zeta } )$
 and 
 $Y_{1m} (\zeta ,\bar{\zeta } )$
 spherical harmonics and we will denote it 
by 
 $V_{I} (u,\zeta ,\bar{\zeta } )$
. The line element becomes
\begin{equation} \label{pr4}
ds_{0}^{2} =\left( -2\frac{\dot{V} _{I} }{V_{I} } r+1\right) du^{2}
+2\;du\;dr-\frac{r^{2} }{V_{I}^{2} P_{0}^{2} } d\zeta \;d\bar{\zeta }.
\end{equation}

It is a well-known result (see for example ref. \cite{Frolov79}) that under 
these conditions the vector $l$ is singular on a timelike world-line $\gamma (\tau )$, which can be determined by the coordinate condition $r=0$.

\section{Second example: massive `particle' with arbitrary motion}

Let us now take the previous line element, associated to the 
timelike curve  $\gamma (\tau )$, and add a term of the form 
 $\frac{-2m}{r} l_{a}l_{b} $
; that is
\begin{equation} \label{pr5}
 ds^{2} =ds_{0}^{2} -\frac{2m}{r} \;du^{2};
\end{equation}
which can also be thought of as a Robinson-Trautman geometry 
where V is restricted to the condition 
 $\eth^2 V=0$. We use Latin indices 
to denote abstract indices. Under this condition we distinguish 
between two cases:

\emph{Geodesic motion:} When the curve 
 $\gamma (\tau )$
 is a geodesic, the resulting 
line element is the Schwarzschild metric, written in a different 
coordinate system.

\emph{Non-geodesic motion:} If the curve 
 $\gamma (\tau )$
 is not a geodesic, the Ricci 
tensor is different from zero. From the above equation one can 
deduce that
\begin{equation} \label{pr6}
R_{ab} =\frac{6m\dot V}{Vr^2}l_al_b.
\end{equation}

We recognize this as the case of the `photon rocket'\cite{Kinnersley69}.

\section{Radiation in the general case}

The general Robinson-Trautman geometry is asymptotically flat. 
This can be seen from the fact that the coordinate r is proportional 
to the luminosity distance, and therefore the conformal factor 

 $\Omega $
, used in the discussion of the asymptotic behavior, can be taken 
just as 
 $r^{-1} $
; then it is deduced from references \cite{Moreschi87} and \cite{Frolov76} that 
the Weyl tensor behaves as  $\Omega $, in the vicinity of future null infinity, 
and from equation (\ref{pr3}) that the Ricci tensor behaves respectively 
as 
 $\Omega ^{2} $
; which implies that the Riemann tensor goes to zero as 
 $\Omega $
 as 
one approaches future null infinity, assuring the asymptotic 
flatness behavior\cite{Moreschi87} of the geometry. This picture could be generalized 
to consider the case of singular behavior for the scalar V on 
future null infinity; to admit the case, for example, of a photon 
rocket reaching this region; this would force us to consider 
the notion of asymptotic flatness with future null infinities 
which are not complete. This generalization, however, does not 
introduce any relevant issue for the following discussion.

It has been proved\cite{Moreschi87} that the most general class of asymptotically 
flat spacetimes have the BMS structure at future null infinity. 
The BMS symmetry group has an infinite dimensional Lie algebra 
with a four-dimensional normal subgroup\cite{Sachs62}, the so called translation 
subgroup. This fact permits one to construct unambiguously the 
Bondi momentum in regular\footnote{By a regular asymptotically flat spacetime we mean 
an at least twice differentiable conformal structure at future 
null infinity.}
 spacetimes. It has also been proved that any asymptotically 
flat spacetime\cite{Moreschi87} shows the radiation behavior found in regular 
spacetimes; that is the radiation content of an asymptotically 
flat spacetime is also an unambiguous concept.

One can further prove that in a regular asymptotically flat spacetime, 
the radiation content is associated with the time variation of 
the Bondi momentum. More concretely, the Bondi momentum can be 
given in terms of the GHP formalism by
\begin{equation} \label{pr7}
P^{\alpha } =-\frac{1}{4\pi } \int \left( \Psi _{2}^{0} +\sigma ^{0}
\frac{\partial \bar{\sigma } ^{0} }{\partial u_{B} } \right)  \hat{l}
^{\alpha } \;dS^{2} =\frac{1}{4\pi } \int \frac{m}{V^{3} }  \hat{l}
^{\alpha } \;dS^{2}
\end{equation}
where we are using a Bondi frame adapted to the Bondi coordinates $(u_{B} ,\zeta ,\bar{\zeta } )$, $\alpha =0,1,2,3$ and
\[
\left( \hat{l} ^{\alpha } \right) =\left( \sqrt{4\pi } \,Y_{00}
,\;-\sqrt{\frac{2\pi }{3} } \;\left( Y_{11} -Y_{1-1} \right)
,\;i\sqrt{\frac{2\pi }{3} } \left( Y_{11} +Y_{1-1} \right)
,\;\sqrt{\frac{4\pi }{3} } \;Y_{10} \right)
.
\]

The Weyl radiation behavior is encoded in the asymptotic components $\Psi _{4}^{0}$ and $\Psi _{3}^{0}$
, which in turn can be expressed in terms in a the Bondi 
frame by its shear, i.e.: $\Psi_4^0 =-\partial^2 \bar \sigma^0 /\partial u_B^2$ and $\Psi_3^0 =-\eth ( \partial \bar \sigma^0
/\partial u_B) $
. The Ricci radiation behavior 
is in turn encoded in the asymptotic component 
 $\Phi _{22}^{0} $.

The time variation of the Bondi momentum in terms of the Bondi 
time is given by
\begin{equation} \label{pr8}
\frac{dP^{\alpha } }{du_{B} } =-\frac{1}{4\pi } \int \left(
\frac{\partial \sigma ^{0} }{\partial u_{B} } \frac{\partial \bar{\sigma }
^{0} }{\partial u_{B} } -\Phi _{22}^{0} \right) \hat{l} ^{\alpha } 
\;dS^{2};
\end{equation}
while the variation of the Bondi momentum in terms of the Robinson-Trautman 
time is
\begin{equation} \label{pr9}
\frac{dP^{\alpha } }{du} =-\frac{1}{4\pi } \int \left( \frac{\eth^{2}V\bar \eth^2 V}{V} -\frac{\Phi
_{22}^{\left( RT\right) 0} }{V^{3} } \right) \hat{l} ^{\alpha }  \;dS^{2}
\end{equation}

In the present case of the Robinson-Trautman geometry the Bondi 
shear is given by
\[
\frac{\partial \sigma ^{0} }{\partial u_{B} } =\frac{\eth^2 V}{V}
\]
and the Ricci 
 $\Phi _{22}^{0} $
 component is given by
\[
\Phi _{22}^{0} =\frac{\Phi _{22}^{RT} }{V^{4} }.
\]

Note that there is no contradiction among the last two equations 
and (\ref{pr8}) and (\ref{pr9}), since $\partial u_{B}/\partial u =V$.

\section{The photon rockets}

A family of axisymmetric metrics corresponding to a particle 
emitting null fluid anisotropically, and therefore accelerating, 
the so called photon rockets, first described by Kinnersley\cite{Kinnersley69}, 
has been recently reanalyzed in reference \cite{Bonnor94}. These metrics 
actually coincide with the line element appearing in equation 
(\ref{pr5}). Using this Kerr-Schild decomposition, the corresponding 
energy momentum tensor was studied in reference \cite{Bonnor94} to estimate 
the matter flux(there called radiation), over a local two-sphere. 
It can be seen from the above presentation that the unambiguous 
radiation is related to the change of momentum by equation (\ref{pr8}) 
which under this circumstances adopts the form
\[
\frac{dP^{\alpha } }{du_{B} } =\frac{1}{4\pi } \int \Phi _{22}^{0} 
\hat{l} ^{\alpha } \;dS^{2} =-\frac{1}{4\pi } \int \frac{3\;m\;\dot{V}
}{V^{5} }  \hat{l} ^{\alpha } \;dS
\]
since in this case $\partial \sigma ^{0} /\partial u_{B}=0$
 due to the fact that $\eth^2 V =0$. In other words, this 
spacetime does not show gravitational radiation, i.e. Weyl radiation, 
and there is only matter radiation.

In reference \cite{Damour95} the photon rocket spacetimes where studied from 
the perspective of linearized gravity. The problem could be stated 
in the following terms; find the linearized solution to Einstein 
equation with the source
\begin{equation} \label{pr10}
T_{ab} =\frac{w(u,\zeta ,\bar{\zeta } )}{4\pi \;r^{2} } \;l_{a} \;l_{b}
\end{equation}
where\footnote{They use the Greek letter epsilon instead of w in that 
reference.} $l_{a} $ is the generator of a congruence of null geodesics emanating 
from $r=0$. More generally, the energy momentum tensor of equation 
(\ref{pr10}) can naturally be considered the source for the generalized 
photon rockets. Equation (3.12) of reference \cite{Damour95} establishes 
that the source must satisfy the following equation of motion:
\[
\frac{d\left( M(u)\;u^{a} \right) }{du} \equiv \frac{d\left( P^{a}
\right) }{du} =-\frac{1}{4\pi } \;\int w\;l^{a} d\Omega  
\]
which can be seen to agree with our equation (\ref{pr9}) above by 
recognizing that the vector $k^a$ of reference \cite{Damour95} is identified with 
our vector $-l^a$; and that $w=-\Phi _{22}^{(RT)} $
; $l^{a} =\hat{l}^{a}/V $ and 
 $d\Omega =dS^{2}/V^{2} $
; since $\eth^{2}V=0$ for the photon rockets. 
Note that a relation between $M(u)$ and $m$ is easily obtained by 
noting that 
 $M(u)^{2} =P^{\alpha } \;P_{\alpha } $
 with 
 $P^{\alpha } $
 given by eq. (\ref{pr7}) above. It is worth while 
to note that although we have chosen $m$ to be a constant, by reparametrizing 
the coordinate $u$, we could have taken any other choice, as for 
example the total (Bondi) mass at the retarded time $u$, as used 
in reference \cite{Damour95}. Our choice simplifies the reading of eq. (\ref{pr2}), 
which would otherwise include another term involving the time 
derivative of $m$.

It is important to emphasize that while the last equation was 
deduced from the local properties of the source, eq. (\ref{pr9}) was 
obtained from the asymptotic structure. It happens that in this 
particular case there is a clear map from inner data to asymptotic 
data.

The information contained in eq. (3.12) of reference \cite{Damour95} coincides 
with that obtained from eqs. (30) and (31) of reference \cite{Bonnor94}.

The linearized solutions for the case of a general source 
 $w(u,\zeta ,\bar{\zeta } )$
 involve 
instead very complicated expressions, as for example equation 
(4.35) of reference \cite{Damour95}. We would like to point out here that 
the problem of the generalized photon rockets (i.e. with the 
source (\ref{pr10})) has exact solutions which are the Robinson-Trautman 
geometries. This can be seen by comparing the Ricci tensor expressed 
by the component shown in eq. (\ref{pr3}) with the energy tensor of eq. 
(\ref{pr10}), from which it is deduced that an exact solution is recognized 
if one identifies $w$ with
\begin{equation} \label{pr11}
w(u,\zeta ,\bar{\zeta } )=3m\frac{\dot{V}}{V} +V^{3}\eth^2 \bar \eth^2 V- V^2 \eth^2 V \bar\eth^2 V.
\end{equation}

In particular, the photon rocket geometries are those for which 
$w$ is constructed with $V$ satisfying 
 $\eth^2V=0$
. Note that equation (\ref{pr11}) 
can be read in two ways: one in which 
 $V(u,\zeta ,\bar{\zeta } )$
 is given and one computes  $w(u,\zeta ,\bar{\zeta } )$
, thus providing an exact solution for the source (\ref{pr10}); and the 
other in which the scalar $w$ is given and one has to look for 
the corresponding $V$. The latter is a problem whose solution is 
outside the scope of the present article.

The Robinson-Trautman geometries were also studied in reference 
\cite{Frolov76} from the point of view of their Petrov classification.

\section*{Acknowledgments}
 
We would like to thank the referees for pointing out several
improvements. This research was supported in part by grants from
CONICET, CONICOR and SeCyT-UNC.


\end{document}